\documentstyle[aps,pre,epsf,epsfig,floats,twocolumn]{revtex} 

\begin{document}
\normalsize
\renewcommand{\textfraction}{0.1}
\bibliographystyle{unsrt} 


\title{Interaction potentials for soft and hard ellipsoids}
\author{
R. Everaers\thanks{present address: Max-Planck-Institut f\"ur Physik komplexer
Systeme, N\"othnitzer Str. 38, 01187 Dresden, Germany} and 
M.R. Ejtehadi\thanks{present address: Department of Physics \& Astronomy, 
                          University of British Columbia, 
                          6224 Agricultural Road, 
                          Vancouver, B.C. V6T 1Z1, 
                          Canada}}
\address{Max-Planck-Institut f\"{u}r Polymerforschung, Postfach 3148,
  D-55021 Mainz, Germany}  

                   \date{\today}
\maketitle
\begin{abstract}
  Using results from colloid science we derive interaction potentials
  for computer simulations of mixtures of soft or hard ellipsoids of
  arbitrary shape and size. Our results are in many respects
  reminicent of potentials of the Gay-Berne type but have a
  well-defined microscopic interpretation and no adjustable parameters.
  Since our potentials require the calculation of similar variables,
  the modification of existing simulation codes for Gay-Berne
  potentials is straightforward. The computational performance should
  remain unaffected.
\end{abstract} 
             
\section{Introduction}
In molecular simulations~\cite{AllenTildesley,FrenkelSmit} short-range
attractive and repulsive interactions are typically represented using
Lennard-Jones (LJ) 6-12 potentials:

\begin{equation}\label{eq:ULJ} 
U_{\mathrm LJ} = 4\epsilon_{\mathrm LJ} 
\left( \left(\frac\sigma r\right)^{12} - 
       \left(\frac\sigma r\right)^{6}
\strut\right) 
\end{equation} 
where $\sigma/r$ is the dimensionless ratio of the effective
particle diameter and the interparticle distance.  While the $r^{-6}$
part has a physical origin in dispersion or van der
Waals-interactions, the $r^{-12}$ repulsion is chosen by mathematical
convenience. For large molecules the evaluation of the interaction
potential involves a computationally expensive double summation of
Eq.~(\ref{eq:ULJ}) over the respective (atomic) interaction sites

\begin{equation}\label{eq:UMacroSum}
U = \sum_{i\in{\mathrm Body\, 1}}\sum_{j\in{\mathrm Body\, 2}} 
U_{\mathrm LJ}(r_{ij})
\end{equation} 
or the evaluation of a double integral

\begin{equation}\label{eq:UMacro}
U = \int_{\mathrm Body\, 1}\int_{\mathrm Body\, 2} 
\rho_1(\vec r) \rho_2({\vec r}^{\ \prime}) U_{\mathrm LJ}(|\vec r-{\vec r}^{\ \prime}|)\ dV\,dV^\prime
\end{equation}
in the corresponding continuum approximation for bodies with simple
geometric shapes and number densities $\rho_i(\vec r)$ of interaction sites.
We will refer to interaction energies obtained by (numerically)
evaluating Eq.~(\ref{eq:UMacro}) as the ``Hamaker'' potential.

As an alternative Gay and Berne~\cite{GB} (GB) proposed the use of
more complicated single-site interaction potentials for rigid
molecules. Their approach is based on a heuristic modification of a
Gaussian overlap potential. While GB potentials provide a
computationally efficient way to introduce anisotropic interactions in
numerical studies of liquid crystalline systems~\cite{Allen99,Miguel02}, they
have frequently been criticised for their unclear microscopic
interpretation~\cite{Perram_pre_96}.  In the present paper, we use
results from colloid science~\cite{Hunter} to derive approximate
interaction potentials for mixtures of ellipsoids of arbitrary size
and shape which have a well-defined microscopic interpretation and
no adjustable parameters.  

The paper is organized as follows: After introducing the Gay-Berne potential
in section~\ref{sec:GB}, we review in section~\ref{sec:Hamaker} the Hamaker
theory for two spheres of arbitrary size and develop a relatively simple
approximation of the interaction potential which is valid at arbitrary
distances. In section~\ref{sec:Deryaguin2infty} we generalize this expression
to the case of interacting ellipsoids. Section~\ref{sec:PoleContacts} presents
a numerical test of this approximation for the case of pole contacts between
aligned ellipsoids. In section~\ref{sec:Relation2GB} we suggest computable
expressions for the orientational dependence of the interaction potential
which are in many respects reminicent of those familiar from GB potentials.  A
numerical test of the proposed interaction potential for ellipsoidal particles
of different shape at arbitrary relative position and orientation is presented
in in section~\ref{sec:NumericalTest}. We conclude with a brief summary in
section~\ref{sec:Summary}.

\section{The Gay-Berne Potential}
\label{sec:GB}

A rigid body $i$ is specified by its center position $\vec r_i$, its
orientation (expressed, for example, via a rotation matrix ${\bf A}_i$
for the transformation from the lab frame to the body frame) and its
shape. In the case of ellipsoids, the shape is given by three
radii $a_i,b_i,c_i$ which can be used to define a ``structure matrix''

\begin{equation}
\label{eq:S}
{\bf S}_i = \left( \begin{tabular}{ccc} 
               $a_i$& 0 &0\\ 
               0&$b_i$ &0 \\ 
               0& 0 &$c_i$  
    \end{tabular} \right) 
\end{equation}
in the body frame of the ellipsoid.

The most general form of a Gay-Berne potential for dissimilar biaxial
ellipsoids was introduced by Berardi, Fava and Zannoni 
(BFZ)~\cite{Berardi95} as a product of three terms:
\begin{eqnarray}
  \label{eq:GB}
  U({\bf A}_1,{\bf A}_2, {\vec r}_{1 2}) &=&   
  U_{\mathrm r}({\bf A}_1,{\bf A}_2, {\vec    r}_{1 2}) \nonumber \\ 
 &\times&  \eta_{1 2} ({\bf A}_1,   {\bf A}_2)  
    \ \    \chi_{1 2} ({\bf A}_1,{\bf A}_2,{\hat{r}}_{1 2})
\end{eqnarray}
where (${\hat{r}}_{1 2}$) ${\vec r}_{1 2}$ is the (unit) vector
between the center positions:
${\vec r}_{1 2}\equiv {\vec r}_{2}-{\vec r}_{1}$ and 
${\hat{r}}_{1 2}\equiv {\vec r}_{1 2}/|{\vec r}_{1 2}|$.

The first term  controls the
distance dependence of the interaction and has the form of a simple LJ
potential Eq~(\ref{eq:ULJ})

\begin{equation}\label{eq:U0} 
U_{\mathrm r} = 4\epsilon_{\mathrm GB} 
\left( \left(\frac\sigma {h_{1 2}+\gamma\sigma}\right)^{12} - 
       \left(\frac\sigma {h_{1 2}+\gamma\sigma}\right)^{6}
\strut\right) 
\end{equation} 
where the interparticle distance $r_{1 2}$ is replaced by the distance 
$h_{1 2}$ of closest approach between the two bodies:

\begin{equation}\label{eq:h}
h_{1 2} \equiv \min(|\vec r_i-\vec r_j|) \ \forall (i\in{\mathrm Body\, 1},j\in{\mathrm Body\, 2}).
\end{equation}
The position of the potential minimum, 
$(2^{1/6}-\gamma)\sigma$ 
is shifted empirically relative to the
Lennard-Jones value $2^{1/6}\sigma$.  Typically $\gamma=1$.  The well depth
is $\min(U_{\mathrm r})=\epsilon_{\mathrm GB}$.

In general, the calculation of $h_{1 2}$ is non-trivial. For ellipsoids a
suitable scheme was worked out by Perram et
al.~\cite{Perram_pre_96,Perram_jcompphys_85,Perram_cpl_84}.
These authors also clarified the meaning of the distance

\begin{eqnarray}
h_{1 2}^{\mathrm GB}({\bf A}_1,{\bf A}_2, {\vec r}_{1 2}) &=& 
  r_{1 2} -\sigma_{1 2}({\bf A}_1,{\bf A}_2, \hat{r}_{1 2}) 
\label{eq:h_GB}\\
\sigma_{1 2}({\bf A}_1,{\bf A}_2, \hat{r}_{1 2}) &=& [{1 \over 2}
 \hat{r}_{1 2}^T\ \  {\bf G}_{1 2}^{-1}({\bf A}_1, {\bf A}_2) \ \
 \hat{r}_{1 2}]^{-1/2}
\label{eq:sigma_12}\\
{\bf G}_{1 2}({\bf A}_1, {\bf A}_2) &=&{\bf A}_1^T {\bf S}_1^2  {\bf A}_1 + {\bf A}_2^T {\bf S}_2^2  {\bf A}_2
\label{eq:G_GB}
\end{eqnarray}
which is usually employed together with the Gay-Berne potential~\cite{Allen96}.
Eq.~(\ref{eq:h_GB}) is an approximation which fails, for example, in
the case of two spheres with unequal radii $a_1\ll a_2$ where
$\sigma_{1 2} = \sqrt{2(a_1^2+a_2^2)}\approx \sqrt{2}a_2\gg a_1+a_2$.
In this article we always use the correct contact distance $h_{1 2}$.
Fig.~\ref{fig:ucomparison}a and b provide a comparison of the quality
of the various approximation schemes, if $h_{1 2}$ is replaced by the
Gay-Berne approximation $h_{1 2}^{\mathrm GB}$.

The two other terms in Eq.~(\ref{eq:GB}) control the interaction strength as a
function of the relative orientation and position of the ellipsoids. The
second term~\cite{symmetry} introduces an empirical exponent $\nu$:

\begin{eqnarray}
\label{eq:eta_GB}
 \eta_{1 2} ({\bf A}_1, {\bf A}_2) &=& 
\left[ \frac{2\, s_1 s_2}{\det[{\bf G}_{1
      2}({\bf A}_1, {\bf A}_2)]} \right]^{\nu/2}\\
\label{eq:eta_numerator_GB}
 s_i &=& [a_i b_i + c_i c_i] [a_i b_i ]^{1/2}
\end{eqnarray}
The third term has the form
\begin{equation}
\label{eq:chi_GB}
 \chi_{1 2}({\bf A}_1,{\bf A}_2,  \hat{r}_{1 2}) = [2 \hat{r}_{1 2}^T\ \  {\bf B}_{1 2}^{-1}({\bf A}_1, {\bf A}_2)  \ \hat{r}_{1 2}]^\mu.
\end{equation}
with
\begin{equation}
\label{eq:B}
  {\bf B}_{1 2}({\bf A}_1, {\bf A}_2) =  {\bf A}_1^T {\bf E}_1  {\bf
    A}_1 + {\bf A}_2^T {\bf E}_2  {\bf A}_2  
\end{equation}
and
\begin{equation}
\label{eq:E}
E_i = \left( \begin{tabular}{ccc} 
               $e_{ai}^{-1/\mu}$& 0 &0\\ 
               0&$e_{bi}^{-1/\mu}$ &0 \\ 
               0& 0 &$e_{ci}^{-1/\mu}$  
    \end{tabular} \right) 
\end{equation}
where $e_{ai},e_{bi},e_{ci}$ characterize the relative well depth for
side-to-side, face-to-face and end-to-end interactions between two
ellipsoids of type $i$.  $\mu$ is another empirical exponent.

To summarize, the physical problem of a mixture of colloidal particles of
equal composition, but of different sizes and (ellipsoidal) shapes, is defined
via Eqs.~(\ref{eq:ULJ}) and (\ref{eq:UMacro}). It requires the specification
of a material constant and of the shapes $a_i,b_i,c_i$ of the involved
particles. Gay-Berne potentials introduce additional adjustable parameters:
the shift parameter $\gamma$, the empirical exponents $\nu$ and $\mu$, and
three energy parameters per particle type. These parameters are usually
adjusted by fitting Eq.~(\ref{eq:GB}) to the numerical evaluation of
Eq.~(\ref{eq:UMacroSum}) for small assemblies of suitably arranged
Lennard-Jones particles~\cite{GB} or specific organic
molecules~\cite{Berardi98}. Note, that there are no additional parameters
specifying the interactions between ellipsoids of {\em different}
shape. Rather, Eqs.~(\ref{eq:U0}) to (\ref{eq:E}) provide heuristic ``mixing
rules'' for this case.

In the following we will partially justify the orientation
dependent part of the Gay-Berne potential
and the implicit mixing rules for {\em particular} choices of the adjustable
parameters. Note, however, that the product ansatz of
Eqs.~(\ref{eq:GB}) incorrectly reduces the attractive and repulsive
parts of the interaction between extended objects at {\em arbitrary}
distances to {\em simple} power laws with distance {\em independent},
shape and orientation {\em dependent} prefactors.  To overcome this
problem we will abandon the strategy initiated by Gay and Berne who
sought modifications to the Lennard-Jones potential for {\em point}
particles. Instead we will try to preserve the case of interacting
spheres of finite volume as a proper limit.

\section{Hamaker theory}
\label{sec:Hamaker}

\begin{figure}[tbp]
  \begin{center}
   \hspace{0cm}
    \psfig{figure=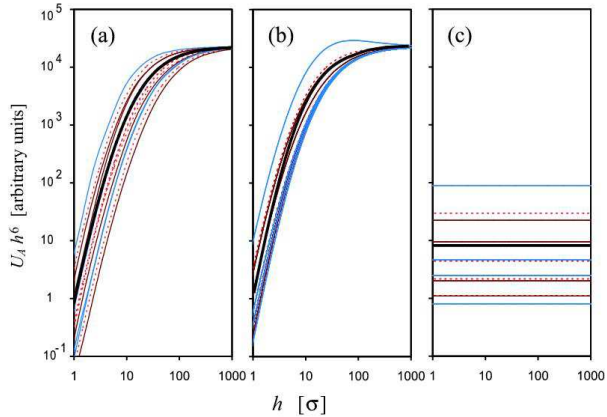,width=8cm}
    \vspace{.3cm}

    \caption{Attractive part $U_A$ of the potential energie multiplied by
    the inverse asymptotic distance dependence $h^6$ as a
    function of the distance of closest approach $h$ for pole contacts between
    differently shaped ellipsoids: (a) Hamaker potential, (b) the
    approximation proposed in the present paper, and (c) a Gay-Berne 6-12
    potential adjusted to reproduce the energy minima within the Deryaguin
    aproximation. The curves converge at large distances, because we have used
    prolate ($(a,b,c)=(1,6,6)\sigma$), oblate ($(a,b,c)=(2,2,9)\sigma$) and
    spherical ($a=36^\frac{1}{3}\sigma$) ellipsoids of identical volume.  The
    plots contain results for sphere-sphere (thick solid line),
    prolate-prolate (gray lines), oblate-oblate (narrow dark lines), and
    prolate-oblate (dotted lines) contacts.  }

    \label{fig:u6h}
  \end{center}
\end{figure}

Eq.~(\ref{eq:UMacro}) can be solved exactly for two spheres of radius
$a_1\le a_2$, volume $V_i=\frac{4\pi}3 a_i^3$ at a distance
$r_{1 2}=(a_1+a_2)+h_{1 2}$ with $h_{1 2}>0$.  For the attractive part of the
interaction Hamaker~\cite{Hamaker_37} obtained

\begin{eqnarray}
U_A&=& -\frac{A_{12}}6 \left[
  \frac{2a_1 a_2}{r_{1 2}^2-(a_1+a_2)^2} + \frac{2a_1 a_2}{r_{1 2}^2-(a_1-a_2)^2}\right.
\nonumber\\
&&
\left.
  + \ln\left( \frac{r_{1 2}^2-(a_1+a_2)^2}{r_{1 2}^2-(a_1-a_2)^2} \right)
\right]
\label{eq:spheresA}
\end{eqnarray}
where $A_{12}$ is usually referred to as Hamaker's constant. Using LJ
units $A_{12}$ is given by $A_{12}=4\pi^2 \epsilon (\rho \sigma^3)^2$.
Similarly, we found for the repulsive part of the LJ potential:

\begin{eqnarray}
U_R&=&\frac{A_{12}}{37800} \frac{\sigma^6}{r_{1 2}}\left[\strut\strut
\right.\nonumber\\
&&
\ \ \frac{r_{1 2}^2 - 7\,r_{1 2}\,\left( {a_1} + {a_2} \right)  + 
    6\,\left( {{a_1}}^2 + 7\,{a_1}\,{a_2} + {{a_2}}^2 \right) }{{\left
       ( r_{1 2} - {a_1} - {a_2} \right) }^7}
\nonumber\\
&&
+\frac{r_{1 2}^2 + 7\,r_{1 2}\,\left( {a_1} + {a_2} \right)  + 
    6\,\left( {{a_1}}^2 + 7\,{a_1}\,{a_2} + {{a_2}}^2 \right) }{{\left
       ( r_{1 2} + {a_1} + {a_2} \right) }^7}
\nonumber\\
&&
-\frac{r_{1 2}^2 + 7\,r_{1 2}\,\left( {a_1} - {a_2} \right)  + 
      6\,\left( {{a_1}}^2 - 7\,{a_1}\,{a_2} + {{a_2}}^2 \right) }
      {{\left( r_{1 2} + {a_1} - {a_2} \right) }^7} 
\nonumber\\
&&
- \frac{r_{1 2}^2 - 7\,r_{1 2}\,\left( {a_1} - {a_2} \right)  + 
      6\,\left( {{a_1}}^2 - 7\,{a_1}\,{a_2} + {{a_2}}^2 \right) }
      {{\left( r_{1 2} - {a_1} + {a_2} \right) }^7} 
\left.\strut\strut\right] \nonumber \\
\label{eq:spheresR}
\end{eqnarray}
Some insight can be gained by
considering three limiting cases: (i) distances which are smaller than the
(curvature) radii of the spheres, (ii) a small sphere
(i.e. a point particle) at an intermediate distance from a much larger sphere,
and (iii) the large distance limit of a 6-12 Lennard-Jones potential with
appropriately renormalized prefactor:

\begin{eqnarray}
U_A&=&  \left\{ \begin{array}{ll}
   {\large -\frac{A_{12}}{12} \frac{2a_1a_2}{a_1+a_2} \frac1{h_{1 2}}} & 
        \mbox{for $0<h_{1 2}\ll a_1$}\\
    -\frac{A_{12}}{6\pi} V_1 \frac1{h_{1 2}^3} & 
        \mbox{for $a_1\ll h_{1 2}\ll a_2$}\\
    -\frac{A_{12}}{\pi^2} V_1 V_2 \frac1{h_{1 2}^6} & 
        \mbox{for $a_1,a_2\ll h_{1 2}$}
             \end{array}
     \right.
\label{eq:spheresAlimits}\\
U_R&=& \left\{ \begin{array}{ll}
    \frac{A_{12}}{2520} \frac{2a_1a_2}{a_1+a_2} 
            \left(\frac\sigma {h_{1 2}}\right)^6 \frac1{h_{1 2}} & 
        \mbox{for $0<h_{1 2}\ll a_1$}\\
    \frac{A_{12}}{45\pi} V_1 \left(\frac\sigma {h_{1 2}}\right)^6 \frac1{h_{1 2}^3} & 
        \mbox{for $a_1\ll h_{1 2}\ll a_2$}\\
    \frac{A_{12}}{\pi^2} V_1 V_2 \left(\frac\sigma {h_{1 2}}\right)^6 \frac1{h_{1 2}^6} & 
        \mbox{for $a_1,a_2\ll h_{1 2}$}
             \end{array}
     \right.
\label{eq:spheresRlimits}
\end{eqnarray}
Fig.~\ref{fig:u6h}a shows a log-log plot of $U_A h^6$ which illustrates the
deviations from the asymptotic power law at small distances. The figure also
contains numerical result for the attractive part of the Hamaker potential for
pole contacts between prolate and oblate ellipsoids. Qualitatively, the curves
resemble each other.  They converge at large distances, because the particles
were chosen to have identical volumes. In contrast, at small distances the
interaction strongly depends on the relative orientation and position of the
non-spherical particles.

\begin{figure}[tbp]
  \begin{center}
   \hspace{0cm}
    \psfig{figure=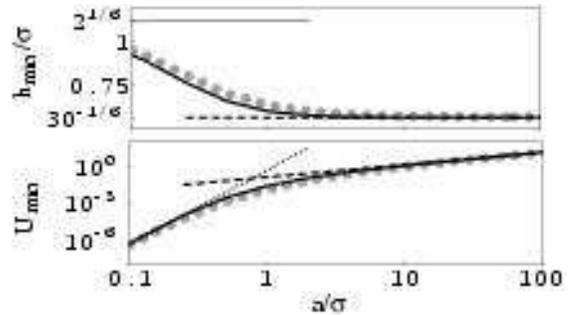,width=8cm}
    \vspace{.3cm}

    \caption{Sphere size dependence of the position (a) and depth (b) of the
    energy minimum: Hamaker potential ($\bullet$)
    Eqs.~(\protect\ref{eq:spheresA}) and (\protect\ref{eq:spheresR}), our
    approximation (------) Eqs.~(\protect\ref{eq:spheresAappr}) and
    (\protect\ref{eq:spheresRappr}), large/Deryaguin (-- --
    --) and small/Lennard-Jones  ($\cdots$) sphere limits
    Eqs.~(\protect\ref{eq:spheresAlimits}) and
    (\protect\ref{eq:spheresRlimits}). } \label{fig:SphereMin} \end{center}
\end{figure}

Since Eqs.~(\ref{eq:spheresA}) and
(\ref{eq:spheresR}) are too complicated for an approximate
generalization, we have instead developed a suitable combination of
the three limiting cases discussed above:

\begin{eqnarray}
U_A &\approx&
  -\frac{A_{1 2}}{36} 
  \left(1+3\,\frac{2a_1\,a_2}{a_1+a_2}\ \frac1{h_{1 2}}\right)
\nonumber\\&&\times
  \left( \frac{a_1}{a_1+{h_{1 2}}/2}\right)^3
  \left( \frac{a_2}{a_2+{h_{1 2}}/2}\right)^3
\label{eq:spheresAappr}\\
U_R &\approx&
  \frac{A_{1 2}}{2025} \left(\frac\sigma h\right)^6
  \left(1+\frac{45}{56}\,\frac{2a_1\,a_2}{a_1+a_2}\ \frac1{h_{1 2}}\right)
\nonumber\\&&\times
  \left( \frac{a_1}{a_1+{h_{1 2}}/60^{1/3}}\right)^3
  \left( \frac{a_2}{a_2+{h_{1 2}}/60^{1/3}}\right)^3
\label{eq:spheresRappr}
\end{eqnarray}
Prefactors were chosen in such a way that Eqs.~(\ref{eq:spheresAappr})
and (\ref{eq:spheresRappr}) reproduce the limits
Eqs.~(\ref{eq:spheresAlimits}) and (\ref{eq:spheresRlimits}) of the
exact results. As a consequence, the approximation is fairly reliable
on all length scales (compare, for example, the thick solid lines in
Figs.~\ref{fig:u6h}a and b). This is also demonstrated in
Fig.~\ref{fig:SphereMin} where we show the sphere size dependence of
the depth and position of the minimum of $U_R+U_A$. While the
short distance expansion becomes reliable for sphere radii $a>5\sigma$,
Eqs.~(\ref{eq:spheresAappr}) and (\ref{eq:spheresRappr}) essentially
reproduce the exact results for arbitrary sphere sizes.

As pointed out above, potentials of the Gay-Berne type Eq.~(\ref{eq:GB})
cannot describe the complex distance dependence of the interaction. But what
about the potential minimum? Within the short distance/large sphere expansion,
its depth $U_{min}$ and position $h_{1 2,min}$ are given by:

\begin{eqnarray}
U_{min} &=& - \frac{30^{1/6}}{14}\ A_{12}\ \frac{2a_1a_2}{\sigma(a_1+a_2)}\\
h_{1 2,min} &=& 30^{-1/6}\sigma
\end{eqnarray}
For the Gay-Berne potential, on the other hand, 

\begin{eqnarray}
U_{min}^{\mathrm GB} &=&  \epsilon_{\mathrm GB}\,\chi_{1 2}\,\eta_{1 2}\\
\chi_{1 2} &=& \frac{e_1 e_2}{\left( \frac12 (e_1^{1/\mu}+e_2^{1/\mu})\right)^\mu}\\
\eta_{1 2} &=& 1\\
h_{1 2,min}^{\mathrm GB} &=& (2^{1/6}-\gamma)\sigma
\end{eqnarray} 
A shift term with 

\begin{equation}\label{eq:gamma}
\gamma = 2^{1/6}-30^{-1/6}\approx 0.56
\end{equation}
in the distances in Eq.~(\ref{eq:U0}) is thus a natural consequence of
the insistence on a 6-12 potential. Furthermore, the comparison suggests
a relation between the energy scale of the GB potential and the
Hamaker constant

\begin{eqnarray}\label{eq:epsilonGB}
\epsilon_{\mathrm GB} &=& A_{12} \frac{30^{1/6}}{14}
\end{eqnarray}
as well as the choices $\mu=1$ and $e_i=a_i/\sigma$. Note, that this
result represents a first justification for one of the empirical GB
mixing rules.

\label{sec:PoleContacts}
\begin{figure*}[t]
  \begin{center} 
   \hspace{0cm} 
   \psfig{figure=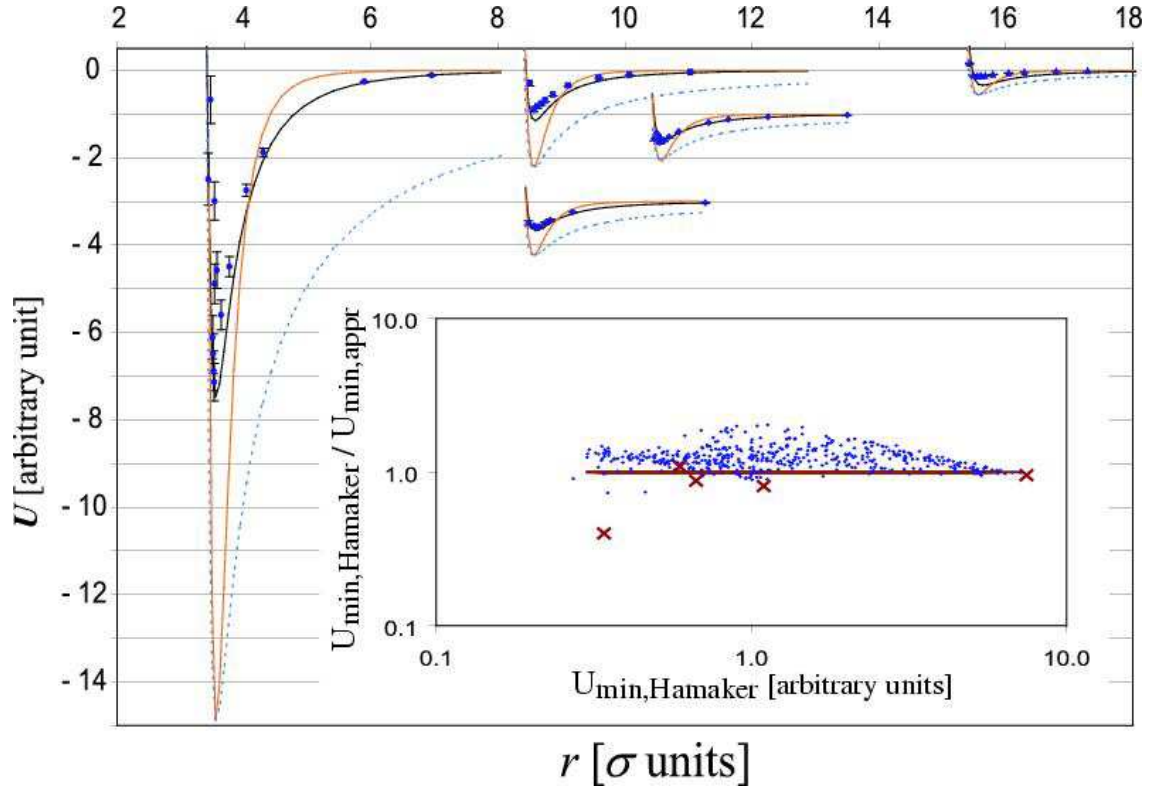,width=15cm}
   \vspace{.3cm} 

   \caption{ Distance dependence of the potential energy for all possible pole
   contacts between an oblate and a prolate ellipsoid with semi-axes
   $(1,6,6)\sigma$ and $(2,2,9)\sigma$ respectively: ($\bullet$) Hamaker
   potential obtained via a Monte Carlo evaluation of the six dimensional
   integral Eq.~(\protect\ref{eq:UMacro}); (-- -- --) Deryaguin approximation
   Eqs.~(\protect\ref{eq:DeryaguinWhite}) and (\protect\ref{eq:DWpoles});
   ($\cdots$) a GB 6-12 potential Eq.~(\protect\ref{eq:GB}) adjusted to
   reproduce the position and depth of the minima in the Deryaguin
   approximation Eqs.~(\protect\ref{eq:gamma}) and
   (\protect\ref{eq:epsilonGB}); (------) our proposal
   Eqs.~(\protect\ref{eq:ellipsoidsAappr}) and
   (\protect\ref{eq:ellipsoidsRappr}). Two data sets were shifted along the
   $y$-axis for clarity reasons.
   Note that none of the approximations contains freely adjustable parameters.
\newline
   The inset shows the ratio of the well depths of the 
   Hamaker potential and of our potential as a
   function of the absolute well depth. ($\times$) Results for pole contacts;
   ($\cdot$) results for ellipsoids with randomly chosen relative orientations
   and positions whose distance is varied along the center-to-center
   line. A comparison of the two data sets allows an evaluation of the quality
   of our approximations for $\chi_{1 2}$ and $\eta_{1 2}$ (see
   Section~\protect\ref{sec:Relation2GB}).}
    \label{fig:minima}
  \end{center}
\end{figure*}

\section{Interpolating between the Deryaguin approximation and the large
distance limit}
\label{sec:Deryaguin2infty}

In the general case of ellipsoids of arbitrary shape, relative position and
orientation, the small and the large distance limit remain (qualitatively)
unchanged. However, for strongly non-spherical ellipsoids with $a_i\ll c_i$
there are new regimes for intermediate distance $a_i\ll h\ll c_i$. For example,
one finds for two thin prolates with semiaxes $(a,a,L)$, $a\ll L$
different power law behavior for parallel, perpendicular, and aligned
configurations:

\begin{eqnarray}
U_{A\,||} &=& \frac{2\pi A_{1 2} }{5} \ \frac{a^4 L}{h^5}
\label{eq:u6par}\\
U_{A\,{\large +}} &=& \frac{\pi A_{1 2} }{2} \ \frac{a^4}{h^4}
\label{eq:u6perp}\\
U_{A\,--} &=& \frac{ A_{1 2} }{30} \ \frac{a^4}{h^2 L^2}
\label{eq:u6alnd}
\end{eqnarray}
We have found no truly satisfactory approximation that would reproduce all
intermediate limiting cases for ellipsoids of arbitrary shape. Nevertheless
we have made some progress compared to a simple 6-12 potential.
Our somewhat naive strategy for developing an approximation for the general
case is (i) to use the Deryaguin approximation in the short distance limit and
(ii) to treat ellipsoids as spheres of equivalent volume on length scales
which exceed the particle diameters.

In the limit where large particles almost touch, the
relevant distances become small compared to the local radii of
curvature of the bodies. The short-distance expansion of
Eq.~(\ref{eq:UMacro}) in powers of the local curvature radii is known
as Deryaguin approximation~\cite{Hunter,Deryaguin_34}. In the most
general case~\cite{White_83}, each body has two different principal radii
of curvature $R_i$ and $R_i^\prime$ at the point of closest approach.
Furthermore, the principal axes of the two surfaces can be rotated by
an angle $\theta$ relative to each other.  The result of the Deryaguin
approximation for Lennard-Jones interactions can be written in the
form

\begin{eqnarray}\label{eq:DeryaguinWhite}
U_{\mathrm DW}(h_{1 2},\theta) &\equiv& \frac{A_{1 2}}{12} 
  \,\chi_{1 2} \,\eta_{1 2} 
  \left(\frac1{210} \left(\frac\sigma {h_{1 2}}\right)^7-
                    \left(\frac\sigma {h_{1 2}}\right)\right)
\end{eqnarray}
where, in analogy to Eq.~(\ref{eq:GB}), we tentatively identify
the orientation and relative position dependent part with a product of
two terms to be specified below:

\begin{eqnarray}\label{eq:DeryaguinWhiteChiEta}
\lefteqn{\chi_{1 2} \,\eta_{1 2} =}&&\\ 
&&  \frac{2\,\sigma^{-1}}
       {\sqrt{\left( \frac{1}{{R_1}} - \frac{1}{{R_1^\prime}} \right)\,
              \left( \frac{1}{{R_2}} - \frac{1}{{R_2^\prime}} \right)\, 
              {\sin (\theta )}^2 
              +
              \left( \frac{1}{{R_1}} + \frac{1}{{R_2}} \right) \,
              \left( \frac{1}{{R_1^\prime}} + \frac{1}{{R_2^\prime}} \right)
              }}\nonumber
\end{eqnarray}
We note that Eq.~(\ref{eq:DeryaguinWhiteChiEta}) defines together with
Eq.~(\ref{eq:gamma}) and (\ref{eq:epsilonGB}), a {\em parameter-free}
6-12 potential which correctly describes the position and depth
of the energy minima for large colloidal particles.

Our second step is the generalization of Eqs.~(\ref{eq:spheresAappr}) and
(\ref{eq:spheresRappr}) to the interaction between ellipsoids~\cite{integrals}

\begin{eqnarray}
\label{eq:ellipsoidsAappr}
\lefteqn{U_A \approx
  -\frac{A_{1 2}}{36} 
  \left(1+3\,\eta_{1 2}\,\chi_{1 2}\ \frac1{h_{1 2}}\right)}
\\
&&\times
  \left( \frac{a_1}{a_1+{h_{1 2}}/2}\right)
  \left( \frac{b_1}{b_1+{h_{1 2}}/2}\right)
  \left( \frac{c_1}{c_1+{h_{1 2}}/2}\right)
\nonumber\\
&&\times
  \left( \frac{a_2}{a_2+{h_{1 2}}/2}\right)
  \left( \frac{b_2}{b_2+{h_{1 2}}/2}\right)
  \left( \frac{c_2}{c_2+{h_{1 2}}/2}\right)
\nonumber\\
\label{eq:ellipsoidsRappr}
\lefteqn{U_R \approx
  \frac{A_{1 2}}{2025} \left(\frac\sigma {h_{1 2}}\right)^6
  \left(1+\frac{45}{56}\,\eta_{1 2}\chi_{1 2}\ \frac1{h_{1 2}}\right)}
\\
&&\times
  \left( \frac{a_1}{a_1+{h_{1 2}}/60^{1/3}}\right)
  \left( \frac{b_1}{b_1+{h_{1 2}}/60^{1/3}}\right)
  \left( \frac{c_1}{c_1+{h_{1 2}}/60^{1/3}}\right)
\nonumber\\
&&\times
  \left( \frac{a_2}{a_2+{h_{1 2}}/60^{1/3}}\right)
  \left( \frac{b_2}{b_2+{h_{1 2}}/60^{1/3}}\right)
  \left( \frac{c_2}{c_2+{h_{1 2}}/60^{1/3}}\right)
\nonumber
\end{eqnarray}
This ansatz reproduces at least the intermediate
power laws for parallely and perpendicularly
oriented thin prolates quite well:

\begin{eqnarray}
U_{A\,||} &=& \frac{4 A_{1 2} }{3} \ \frac{a^4 L}{h^5}
\label{eq:u6parappr}\\
U_{A\,{\large +}} &=& \frac{4 A_{1 2} }{9} \ \frac{a^4}{h^4}
\label{eq:u6perpappr}\\
U_{A\,--} &=& \frac{4 A_{1 2} }{9} \ \frac{a^4}{h^4}
\label{eq:u6alndappr}
\end{eqnarray}
In contrast, our ansatz overestimates the potential at intermediate distances
for needles aligned along their long axes. However, the following comparisons
will show that the deviations are typically within a factor of two or three
{\em over the entire range of distances.}

\section{Pole contacts between biaxial ellipsoids}

The evaluation of Eq.~(\ref{eq:DeryaguinWhiteChiEta}) is
straightforward for pole contacts between aligned ellipsoids with
${\bf A}_1={\bf A}_2$ so that $\theta=0$.  For two biaxial ellipsoids
which touch at their $c$ poles, the principle curvature radii at the
touch point are $R_i = {a_i^2 \over c_i }$ and $R'_i = {b_i^2 \over  c_i}$. 
In this case Eq.~(\ref{eq:DeryaguinWhiteChiEta}) reduces to

\begin{eqnarray}\label{eq:DWpoles}
\chi_{1 2} \,\eta_{1 2} =  
\frac{2\,\sigma^{-1}}
       {\sqrt{\left( \frac{c_1}{{a_1^2}} + \frac{c_2}{{a_2^2}} \right) \,
              \left( \frac{c_1}{{b_1^2}} + \frac{c_2}{{b_2^2}} \right)
              }}
\end{eqnarray}
As in the case of interacting spheres, we can check the quality of our
approximations by comparing to results obtained by, in the present
case numerically, integrating Eq.~(\ref{eq:UMacro}) for pole contacts
between oblate and prolate ellipsoids. Fig.~\ref{fig:minima} shows
good agreement for the shape, depth, and position of the minima.  In
particular, the figure demonstrates that for small molecules
Eqs.~(\ref{eq:ellipsoidsAappr}) and (\ref{eq:ellipsoidsRappr}) provide
a significant improvement over the Deryaguin approximation (or a 6-12
potential taylored to reproduce the minima of the Deryaguin
approximation). Figs.~\ref{fig:u6h}a and b show that we also reproduce
the crossover to the asymptotic behavior at large distances reasonably
well, at least relative to the parameter-free Gay-Berne potential
(Fig.~\ref{fig:u6h}c).

\section{Computable expressions for arbitrary contact geometries and the relation to the Gay-Berne potential}
\label{sec:Relation2GB}
\begin{figure}[t]
  \begin{center}
   \hspace{0cm}
    \psfig{figure=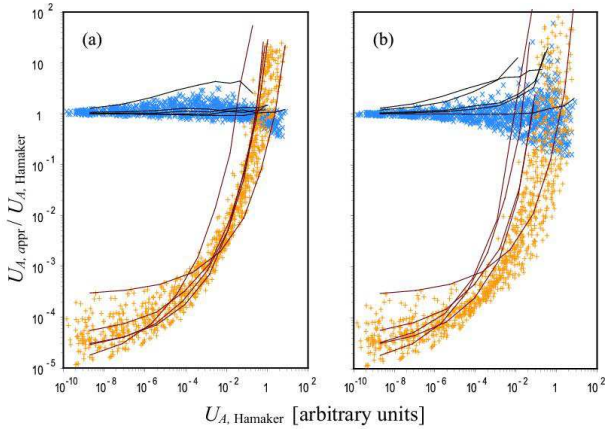,width=8cm}
    \vspace{.3cm}

    \caption{ Deviations of approximate potentials from the true
    Hamaker potential as a function of the absolute value of the Hamaker
    potential: (a) using the correct distance of closest approach
    Eq.~(\ref{eq:h}); (b) using the approximate GB distance function
    Eq.~(\ref{eq:h_GB}). We show results for contacts between an oblate and a
    prolate ellipsoid with semi-axes $(a,b,c)=(1,6,6)\sigma$ and
    $(a,b,c)=(2,2,9)\sigma$ respectively and randomly chosen relative
    orientations and positions: (dark $\times$) our potential, (gray $+$) a
    Gay-Berne 6-12 potential adjusted to reproduce the energy minima within
    the Deryaguin aproximation. The solid lines show the corresponding results
    for pole contacts (see Fig.~\protect\ref{fig:u6h}). }
    \label{fig:ucomparison}
  \end{center}
\end{figure}

In the next step, we take a closer look at the orientation dependent
prefactor in Eq.(\ref{eq:DeryaguinWhite}). We are less interested in
its exact calculation (i.e.  the determination of the contact points,
the local curvature radii and the angle between the principle axis on
the two surfaces) than in finding expressions which offer a good
compromise between computability and correctness. Quite interestingly,
our final expressions turn out to be closely related to those used in
Gay-Berne potentials.

We begin by providing definitions for $\chi_{12}$ and $\eta_{12}$ which,
when multiplied with each other, reproduce Eq.~(\ref{eq:DeryaguinWhiteChiEta})
and whose structure resembles Eqs.~(\ref{eq:chi_GB}) and (\ref{eq:eta_GB}):
\begin{eqnarray}
\chi_{12}&\equiv&\left(\frac{2\, \sigma^{-1}}{\kappa_1^{-1} + 
    \kappa_2^{-1}}\right)
\label{eq:chi_def}\\
\eta_{12}&\equiv&{\kappa_1 + \kappa_2 \over \det[{\bf Q}(\theta)]
  ^{1/2}}
\label{eq:eta_def}\\
\kappa_i &=& \det\left[\left( \begin{tabular}{cc} $R_i$& 0\\ 0& $R'_i$
    \end{tabular} \right) \right]^{1/2}
\label{eq:kappa_det}\\
{\bf Q}(\theta)&\equiv& \left( \begin{tabular}{cc} $R_1$& 0\\ 0& $R'_1$
    \end{tabular} \right) + {\bf \Omega}(\theta)^t    \left( \begin{tabular}{cc} $R_2$& 0\\ 0& $R'_2$
    \end{tabular} \right) {\bf \Omega}(\theta)\nonumber \\
&&\label{eq:Q}
\end{eqnarray}
$\kappa_i \equiv \sqrt{R_i\,R'_i}$ is the Gaussian curvature at the
touch point, ${\bf \Omega}(\theta)$ is a two dimensional rotation
matrix. For two spherical surfaces with $R_1=R'_1$ and $R_2=R'_2$ the
second term reduces to $\eta_{1 2}\equiv1$.

The GB definition Eqs.~(\ref{eq:chi_GB}) - (\ref{eq:E}) of $\chi_{1 2}$
agrees for pole contacts with Eq.~(\ref{eq:chi_def}) {\em provided}
the parameter $\mu$ is set to

\begin{equation}\label{eq:mu_1}
\mu\equiv1
\end{equation}
and the three BFZ energy parameters Eq.~(\ref{eq:E}) are identified
with the Gaussian curvatures at the three poles

\begin{eqnarray}
\label{eq:E_RE2}
{\bf E}_i &=&  \sigma
\left(\begin{array}{ccc} {a_i \over b_i\,c_i} & 0 & 0 \\ 
                         0 & {b_i \over a_i\,c_i} & 0 \\ 
                         0 &0 & {c_i \over a_i\,b_i}
      \end{array}\right)\\
&=& \frac\sigma{\det[{\bf S}_i]} {\bf S}_i^2
\label{eq:E_S_RE2}
\end{eqnarray}

The relation between the GB definition Eq.~(\ref{eq:eta_GB}) of
$\eta_{1 2}$ and Eq.~(\ref{eq:eta_def}) is less direct. The reason is
that the curvature matrices in Eq.~(\ref{eq:Q}) characterize {\em
  surfaces} and, as a consequence, are two dimensional. Similarly, the
angle $\theta$ describes the relative rotation of the two surfaces
around {\em their common normal vector at the points of closest
  approach}. In contrast, GB variables 
characterize the shape and orientation of three dimensional bodies. In
the following, we will present a heuristic combination of GB
variables which reproduces Eq.~(\ref{eq:eta_def}) for pole
contacts.

As a first step, we consider the definition of the Gaussian curvature
in terms of the curvature matrix, Eq.~(\ref{eq:kappa_det}). At the 
$c$-pole $\kappa_{ci}$ can be written in the form

\begin{equation}
\kappa_{ci} 
= \left( \det\left[\left( 
   \begin{tabular}{cc} $\frac{a_i^2}{c_i}$& 0\\ 
                       0& $\frac{b_i^2}{c_i}$
    \end{tabular} \right) \right]\right)^{1/2}
= \left(\frac1{c_i} \det\left[\frac1{c_i} {\bf S}^2_i \right]\right)^{1/2}
\end{equation}
For arbitrary orientations it is tempting to replace $c_i$ by the projected
diameter 

\begin{equation}
\label{eq:sigma_i_RE2}
\sigma_i({\bf A}_i, \hat{r}_{1 2}) \equiv 
\left(\hat{r}_{1 2}^T\ \ {\bf A}_i^T {\bf S}_{i}^{-2}{\bf A}_i \ \
 \hat{r}_{1 2}\right)^{-1/2}
\end{equation}
In this manner we arrive at the following approximative expression for 
$\eta_{1 2}$:

\begin{eqnarray}
\label{eq:eta_RE2}
 \eta_{1 2} ({\bf A}_1, {\bf A}_2) &=& 
   \frac{ \det[{\bf S}_1]/\sigma_1^{2}+
          \det[{\bf S}_2]/\sigma_2^{2}
        }
        {\left(\det[{\bf H}_{12}]/(\sigma_1+\sigma_2)\right)^{1/2}} 
\\
\label{eq:H_RE2}
{\bf H}_{1 2}({\bf A}_1,{\bf A}_2, \hat{r}) &=&
\frac1{\sigma_1}{\bf A}_1^T {\bf S}_1^2  {\bf A}_1 + 
\frac1{\sigma_2}{\bf A}_2^T {\bf S}_2^2  {\bf A}_2
\end{eqnarray}
As a side result we note that the expressions for $\chi_{1 2}$ and
$\eta_{1 2}$ can be further simplified for those contacts which
dominate in the ordered phases of typical liquid crystals, i.e.
contacts between similar poles of identical ellipsoids $(a_1,b_1,c_1)
= (a_2,b_2,c_2) =(a,b,c)$. 

\begin{eqnarray}
\label{eq:chi_sim_RE2}
\chi_{1 2}({\bf A}_1,{\bf A}_2, \hat{r}_{1 2}) &=&   
  \frac{4\sigma^{-1} \det[{\bf S}]}
       {\sigma_{1 2}^2({\bf A}_1,{\bf A}_2, \hat{r}_{1 2})}\\
\label{eq:eta_sim_RE2}
\eta_{1 2}({\bf A}_1,{\bf A}_2) &=&   
  \frac{2^{3/2} \det[{\bf S}]}
       {\left(\det[{\bf G}_{12}]\right)^{1/2}}
\end{eqnarray}
All in all, we are lead to expressions which are in remarkable
agreement with those proposed by BFZ. In the case of $\chi_{1 2}$, we 
employ the same functional form together with particular choices for
the four adjustable parameters $\mu$, $e_{ai}$, $e_{bi}$, $e_{ci}$. 
In the case of $\eta_{1 2}$,
our Eqs.~(\ref{eq:eta_RE2}) and (\ref{eq:H_RE2}) respectively
Eq.~(\ref{eq:eta_sim_RE2}) resemble the corresponding
Eqs.~(\ref{eq:eta_GB}) and (\ref{eq:G_GB}) so strongly, that it seems
clear that our consideration eliminate with $\nu\equiv1$
the last remaining free parameter of the GB potential. 

We note that our proposed modifications leave with $\chi_{1 2}$ and $\eta_{1
2}$ the most CPU time intensive part of the GB potential essentially
unchanged. The small number of additional scalar operations necessary for the
evaluation of Eqs.~(\ref{eq:ellipsoidsAappr}) and (\ref{eq:ellipsoidsRappr})
hardly affects the performance of simulation codes.

\section{Numerical test of the approximations for arbitrary relative
position and orientation of the ellipsoids}
\label{sec:NumericalTest}

The most important question is, of course, how reliable the proposed
approximations are.  Similarly to Eqs.~(\ref{eq:ellipsoidsAappr}) and
(\ref{eq:ellipsoidsRappr}), the combination of Eqs.~(\ref{eq:chi_GB}),
(\ref{eq:mu_1}), (\ref{eq:E_S_RE2}), (\ref{eq:eta_RE2}) and (\ref{eq:H_RE2})
can only be considered as an educated guess for
Eq.~(\ref{eq:DeryaguinWhiteChiEta}). The fact that we reproduce the results of
the Deryaguin approximation for pole contacts inspires some confidence, but
otherwise we have made substancial and uncontrolled approximations which need
to be checked against the numerical evaluation of Eq.~(\ref{eq:UMacro}) for
various relative positions and orientations. We represent the results by
plotting the ratio of the approximative and the exact energy as a function of
the exact energy.  In this manner, results from a high-dimensional parameter
space are (i) projected onto a single axis and (ii) sorted by importance.  The
inset in Fig.~\ref{fig:minima} shows a comparison of the depths of the energy
wells close to contact for fixed random orientations, while
Fig.~\ref{fig:ucomparison}a deals with the attractive part of the interaction
at arbitrary distances.  Both figures also contain results for the pole
contacts discussed before to allow for an independent evaluation of the
quality of the approximations for the distance and for the orientational part
of the interaction potential.

When judged against the corresponding results for the Gay-Berne potential, the
agreement between our proposal and the numerical evaluation of
Eq.~(\ref{eq:UMacro}) is excellent. In absolute terms, the deviations do not
exceed a factor of two to three in either direction. Quite interestingly, our
approximations for $\chi_{1 2}$ and $\eta_{1 2}$ do not seem to be the source
of large additional errors. Fig.~\ref{fig:ucomparison}b shows that the
agreement is significantly reduced, if the approximate GB distance function
Eq.~(\ref{eq:h_GB}) is used instead of the true distance of closest approach
Eq.~(\ref{eq:h}).

\section{Summary}
\label{sec:Summary}
We have presented an approximative interaction potential for soft ellipsoidal
particles. Our potential uses (almost) the same variables as the Berardi, Fava
and Zannoni~\cite{Berardi95} form of the Gay-Berne~\cite{GB} potential for
biaxial ellipsoids, agrees significantly better with the numerically evaluated
exact interaction potential, has no unphysical limits, and avoids the
introduction of empirical adjustable parameters. The main modifications we
propose are

\begin{enumerate}
\item to abandon the unphysical factorization of the
orientation and distance dependent parts of the potential (Eq.~\ref{eq:GB}) as
well as the Lennard-Jones like form of the distance dependence itself
(Eq.~\ref{eq:U0}) and to replace them by Eqs.~(\ref{eq:ellipsoidsAappr}) and
(\ref{eq:ellipsoidsRappr}).
\item to use the Gaussian curvatures at the ellipsoid poles
Eqs.~(\ref{eq:mu_1}) - (\ref{eq:E_S_RE2}) in order to characterize the
relative well depth for side-to-side, face-to-face and end-to-end interactions
through the orientation and relative position dependent factor 
$\chi_{12}({\bf A}_1,{\bf A}_2, \hat{r}_{1 2})$ defined in Eqs.~(\ref{eq:chi_GB}) 
and (\ref{eq:B}).
\item to replace the definition Eqs.~(\ref{eq:G_GB})
- (\ref{eq:eta_numerator_GB}) of the purely orientiation dependent
factor $\eta_{1 2}({\bf A}_1,{\bf A}_2)$
by Eqs.~(\ref{eq:sigma_i_RE2}) - (\ref{eq:H_RE2}).
\item to use the (effective) Hamaker constant $A_{1 2}$ \cite{Hunter} to set
the energy scale.
\end{enumerate}
Our results for the attractive part of the soft potential are directly
applicable to hard ellipsoids with van-der-Waals interactions. Furthermore,
the proposed potential comprises the interaction of point particles with
ellipsoids as a well defined limit. This may be of interest for studies of
wetting or polymer adsorption in colloidal dispersions which so far assume
either a flat or a cylindrical geometry~\cite{MilchevBinder02}.

\section*{Acknowledgements}
We gratefully acknowledge extended discussions with L. Delle Site, M. Allen
and F. M\"uller-Plathe and benefitted from a copy of a Gay-Berne simulation
code provided by M. Allen.  The authors thank the DFG for the financial
support of this work within an Emmy-Noether grant.


\begin{thebibliography}{10}
\bibitem{symmetry}
  There is a slight problem with
  the BFZ suggestion for $\eta_{1 2}$, because the definition of the
  numerator leads to different energies for two physically
  indistinguishable situations: (i) rotation of an ellipsoid by 90
  degrees around its $b$-axis and (ii) exchange of the values of $a_i$
  and $c_i$.
\bibitem{integrals}
The functional form can at least partially be justified by the observation that
the integrations in Eq.~(\ref{eq:UMacro}) are of a type which produces simple
results in two limiting cases:
\begin{eqnarray}
\lefteqn{\int_{-d/2}^{d/2} (r+x)^{-n} dx }&&\nonumber\\ 
&\sim&\left\{ \begin{array}{ll}
    h^{-(n-1)} & \mbox{for $h=r-d/2\ll d$}\\
    d\, h^{-n} & \mbox{for $h\approx r\gg d$}
             \end{array}
     \right.\nonumber\\
&\sim& h^{-(n-1)} \frac d{d+h}\nonumber
\end{eqnarray}
(i) For separations $h$ which are much smaller than the particle
extension the upper limit of the integration can be extended to
infinity, reducing the exponents in the power laws describing the
interaction by one. (ii) For large separations, the functional form of
the interaction is not affected by the integration which only produces
a prefactor of the order of the respective particle diameter.
\end{thebibliography}
\end{document}